\documentclass[a4paper,12pt,epsf]{article}

\usepackage{amsmath}
\usepackage[psamsfonts]{amssymb}
\usepackage{rsfs}
\usepackage{bm}

\usepackage{cite}
\usepackage[dvips]{graphicx}
\usepackage{color}

\begin{document} 

\begin{titlepage}

\baselineskip 10pt
\hrule 
\vskip 5pt
\leftline{}
\leftline{Chiba Univ./KEK Preprint
          \hfill   \small \hbox{\bf CHIBA-EP-158}}
\leftline{\hfill   \small \hbox{\bf KEK Preprint 2006-2}}
\leftline{\hfill   \small \hbox{hep-lat/0604016}}
\leftline{\hfill   \small \hbox{October 2006}}
\vskip 5pt
\baselineskip 14pt
\hrule 
\vskip 0.5cm
\centerline{\Large\bf 
Compact lattice formulation of 
} 
\vskip 0.3cm
\centerline{\Large\bf  
Cho-Faddeev-Niemi decomposition: 
}
\vskip 0.3cm
\centerline{\Large\bf  
string tension from magnetic monopoles 
}
\centerline{\large\bf  
}

\vskip 0.3cm

\centerline{{\bf 
S. Ito$^{\star,{1}}$, 
S. Kato$^{\sharp,{2}}$, 
K.-I. Kondo$^{\dagger,\ddagger,{3}}$,  
T. Murakami$^{\ddagger,{4}}$,
A. Shibata$^{\flat,{5}}$  
$\&$
T. Shinohara$^{\ddagger,{6}}$
}}  
\vskip 0.5cm
\centerline{\it
${}^{\star}$Nagano National College of Technology, 716 Tokuma, Nagano 381-8550, Japan
}
\vskip 0.3cm
\centerline{\it
${}^{\sharp}$Takamatsu National College of Technology, Takamatsu 761-8058, Japan
}
\vskip 0.3cm
\centerline{\it
${}^{\dagger}$Department of Physics, Faculty of Science, 
Chiba University, Chiba 263-8522, Japan
}
\vskip 0.3cm
\centerline{\it
${}^{\ddagger}$Graduate School of Science and Technology, 
Chiba University, Chiba 263-8522, Japan
}
\vskip 0.3cm
\centerline{\it
${}^{\flat}$Computing Research Center, High Energy Accelerator Research Organization (KEK),  
}
\vskip 0.1cm
\centerline{\it
Tsukuba 
305-0801, 
\& 
Graduate Univ. for Advanced Studies (Sokendai),
Japan
}
\vskip 0.1cm
\centerline{\it
}
\vskip 0.3cm
\vskip 0.5cm

\begin{abstract}
In this paper we begin on a new lattice formulation of the non-linear change of variables called the Cho--Faddeev--Niemi decomposition in  SU(2) Yang-Mills theory. 
This is a compact lattice formulation improving the non-compact lattice formulation proposed in our previous paper.  Based on this formulation, we propose a new gauge-invariant definition of the magnetic monopole current which guarantees the magnetic charge quantization and reproduces the conventional magnetic-current density obtained in the Abelian projection based on the DeGrand--Toussaint method. 
Finally, we demonstrate the magnetic monopole dominance in the string tension  in SU(2) Yang-Mills theory on a lattice. 
Our formulation enables one to reproduce in the gauge-invariant way remarkable  results obtained so far only in the Maximally Abelian gauge. 
\end{abstract}


\hrule  
\vskip 0.1cm
${}^1$ 
  E-mail:  {\tt shoichi@ei.nagano-nct.ac.jp}

${}^2$ 
  E-mail:  {\tt kato@takamatsu-nct.ac.jp}
  
${}^3$ 
  E-mail:  {\tt kondok@faculty.chiba-u.jp}
  
${}^4$ 
  E-mail:  {\tt tom@cuphd.nd.chiba-u.ac.jp}
  
${}^5$ 
  E-mail:  {\tt akihiro.shibata@kek.jp}
  
${}^6$ 
  E-mail:  {\tt sinohara@graduate.chiba-u.jp}

\par 
\par\noindent



\vskip 0.5cm

\newpage
\pagenumbering{roman}




\end{titlepage}


\pagenumbering{arabic}

\baselineskip 14pt
\section{Introduction}

In the  previous paper \cite{KKMSSI05}, we have demonstrated that the gauge-invariant magnetic monopole can be constructed in the pure Yang-Mills theory without any fundamental scalar field. 
The success is achieved based on a new viewpoint proposed by three of us \cite{KMS05} for the non-linear change of variables (NLCV), the so-called Cho--Faddeev--Niemi (CFN) decomposition which was first made by Cho \cite{Cho80} and  has recently been readdressed by Faddeev and Niemi \cite{FN98}, see also \cite{Shabanov99}. 
We have for the first time formulated the NLCV on a  lattice  to perform non-perturbative investigations.   Moreover, we have proposed a lattice construction of the gauge-invariant magnetic monopole which we called the lattice CFN monopole \cite{KKMSSI05}. It was shown that the magnetic-current density defined from the lattice CFN monopole reproduces the conventional monopole based on  the DeGrand and Toussaint (DT) method \cite{DT80}  on a lattice.  
However, the previous construction  of the lattice CFN decomposition and the resulting lattice CFN monopole have the following disadvantages. 

\begin{enumerate}
\item
Gauge invariance on a lattice is manifest only to the first order of the lattice spacing $\epsilon$.

\item
The lattice artifact sets in at   first order of the lattice spacing, since the finite difference is used in the calculations of differentiation.

\item
It is not obvious whether the resulting magnetic charge is quantized, since the monopole current is a real-valued variable. 

\end{enumerate}

In this paper, we begin on a new lattice formulation of the NLCV or CFN decomposition, which resolves simultaneously all the issues raised above.  This is nothing but a compact formulation of the Yang-Mills theory in terms of new  lattice variables.  Then we introduce a new definition of gauge-invariant magnetic monopole on a lattice based on the new formulation. 
The new formulation resolves simultaneously the first two issues, because we can define the electric and magnetic fields by  a plaquette variable constructed from the new link variable as an element of a compact gauge group SU(2).   The third issue is resolved and  quantization of the magnetic charge is guaranteed, if  the magnetic monopole is defined through the gauge-invariant flux obtained by a plaquette variable. 
Moreover, we confirm  dominance of the new monopole  in the string tension, while it was first shown in \cite{SNW94} in the conventional Maximally Abelian (MA) gauge \cite{KLSW87}.

\section{Lattice CFN variables or NLCV on a lattice}

In this paper we propose a natural and useful formulation  on a lattice of  the non-linear change of variables (NLCV) corresponding to the CFN decomposition \cite{Cho80,FN98} in the continuum formulation. 
It is a minimum requirement that such a lattice formulation must reproduce  the continuum counterparts in the naive continuum limit. 
In this stage, therefore, it is instructive to recall how the NLCV are defined in the continuum formulation. 
We restrict the following argument to SU(2) gauge group, since we wish to avoid technical difficulties and we give numerical results only for SU(2).%
\footnote{
However, the following argument for SU(2) can be extended to the SU(3) case without many difficulties as will be shown in a subsequent paper.
}

In the continuum formulation \cite{Cho80,KMS05}, a color vector field $\vec{n}(x)=(n_A(x))$ $(A=1,2,3)$ is introduced as a three-dimensional unit vector field. 
In what follows, we use the boldface to express the Lie-algebra $su(2)$-valued field, e.g., ${\bf n}(x) :=n_A(x)T_A$, $T_A=\frac{1}{2}\sigma_A$ with Pauli matrices $\sigma_A$ ($A=1,2,3$).
Then  the $su(2)$-valued gluon field (gauge potential) $\mathbf{A}_\mu(x)$ is decomposed into two parts:
\begin{align}
  \mathbf{A}_\mu(x) = \mathbf{V}_\mu(x) + \mathbf{X}_\mu(x) ,
\end{align}
in such a way that the color vector field ${\bf n}(x)$ is covariantly constant in the background  field $\mathbf{V}_\mu(x)$:
\begin{align}
 0 = \mathscr{D}_\mu[\mathbf{V}] {\bf n}(x) 
:= \partial_\mu {\bf n}(x) -i g [\mathbf{V}_\mu(x) , {\bf n}(x) ],
 \label{covariant-const}
\end{align}
and that the remaining field $\mathbf{X}_\mu(x)$ is perpendicular to ${\bf n}(x)$:
\begin{align}
  \vec{n}(x) \cdot \vec{X}_\mu(x) \equiv 2{\rm tr}({\bf n}(x)   \mathbf{X}_\mu(x)) = 0 .
  \label{nX=0}
\end{align}
Here we have adopted the normalization ${\rm tr}(T_A T_B)= \frac12 \delta_{AB}$. Both 
${\bf n}(x)$ and $\mathbf{A}_\mu(x)$ are Hermitian fields. This is also the case for $ \mathbf{V}_\mu(x)$ and $\mathbf{X}_\mu(x)$.

By solving the defining equation (\ref{covariant-const}), the   $\mathbf{V}_\mu(x)$ field is obtained in the form:
\begin{align}
  \mathbf{V}_\mu(x) 
  = \mathbf{V}_\mu^{\parallel}(x) + \mathbf{V}_\mu^{\perp}(x)
  = c_\mu(x) {\bf n}(x)   -i g^{-1} [ \partial_\mu {\bf n}(x), {\bf n}(x) ] ,
  \label{Vdef}
\end{align}
where the second term 
$
 \mathbf{V}_\mu^{\perp}(x) := -i g^{-1} [ \partial_\mu {\bf n}(x), {\bf n}(x) ]
 = g^{-1} (\partial_\mu \vec{n}(x) \times \vec{n}(x))_A T_A 
$ 
is perpendicular to ${\bf n}(x)$, i.e., 
$
 \vec{n}(x) \cdot \vec{V}_\mu^{\perp}(x) \equiv 2{\rm tr}({\bf n}(x)   \mathbf{V}_\mu^{\perp}(x)) = 0
$.
Here it should be remarked that the parallel part $\mathbf{V}_\mu^{\parallel}(x)=c_\mu(x) {\bf n}(x)$, $c_\mu(x)= {\rm tr}({\bf n}(x)  \mathbf{A}_\mu(x))$ proportional to ${\bf n}(x)$ can not be determined uniquely from the defining equation (\ref{covariant-const}). 

On a lattice, on the other hand, we introduce the site variable ${\bf n}_{x}$ constructed according to \cite{KKMSSI05}, in addition to the original link variable $U_{x,\mu}$ which is related to the gauge potential $\mathbf{A}_\mu(x)$ in a naive way: 
\footnote{
In general, the argument of the exponential in (\ref{def-U}) is the line integral of a gauge potential along a link from $x$ to $x+\mu$.
We adopt this convention in agreement with our previous paper \cite{KKMSSI05}. Note also that we define a color vector field 
${\bf n}(x) :=n_A(x)T_A$ in the continuum, while ${\bf n}_x  :=n^A_x \sigma_A$ on the lattice for convenience.

}
\begin{align}
U_{x,\mu} = \exp( -i \epsilon g \mathbf{A}_\mu(x)) , 
\label{def-U}
\end{align}
where  ${\bf n}_{x}$ is Hermitian, ${\bf n}_{x}^\dagger={\bf n}_{x}$, and $U_{x,\mu}$ is unitary, $U_{x,\mu}^\dagger=U_{x,\mu}^{-1}$.  
The link variable $U_{x,\mu}$ and the site variable ${\bf n}_{x}$ transform under the gauge transformation II \cite{KMS05} as
\begin{align}
  U_{x,\mu} \rightarrow \Omega_{x} U_{x,\mu} \Omega_{x+\mu}^\dagger = U_{x,\mu}' , \quad
  {\bf n}_{x} \rightarrow \Omega_{x} {\bf n}_{x} \Omega_{x}^\dagger = {\bf n}_{x}' .
\end{align}

Suppose we have obtained a link variable $V_{x,\mu}$ as a group element of $G=SU(2)$, which is related to the  $su(2)$-valued background field $\mathbf{V}_\mu(x)$ through 
\begin{align}
  V_{x,\mu} = \exp (-i\epsilon g \mathbf{V}_\mu(x)) ,
\end{align}
where $\mathbf{V}_\mu(x)$ is to be identified with the continuum CFN variable (\ref{Vdef}) and hence  $V_{x,\mu}$ must be unitary $V_{x,\mu}^\dagger=V_{x,\mu}^{-1}$.

A lattice version of (\ref{covariant-const}) and (\ref{nX=0}) is respectively given by
\begin{align}
 {\bf n}_{x} V_{x,\mu}  = V_{x,\mu} {\bf n}_{x+\mu} ,
 \label{Lcc}
\end{align}
and
\begin{equation}
 {\rm tr}({\bf n}_{x} U_{x,\mu} V_{x,\mu}^\dagger) 
  = 0 .
  \label{cond2m}
\end{equation}
Both conditions must be imposed to determine $V_{x,\mu}$ for a given set of ${\bf n}_{x}$ and $U_{x,\mu}$.  
A lattice version of the defining equation (\ref{covariant-const}) needs  a lattice covariant derivative for an adjoint field.  
We adopt a definition of the covariant derivative for {\it arbitrary} background ${\bf V}_\mu(x)$:
\begin{align}
 D_\mu^{(\epsilon)}[\mathbf{V}] {\bf n}_{x} := \epsilon^{-1}[V_{x,\mu} {\bf n}_{x+\mu} - {\bf n}_{x} V_{x,\mu}]  ,  
 \label{cderivative}
\end{align}
by the following reasons. 
i) When $V_{x,\mu} \equiv {\bf 1}$, the  derivative (\ref{cderivative}) reduces to the (forward) lattice derivative 
$
 \partial_\mu^{(\epsilon)} {\bf n}_{x} := \epsilon^{-1}[{\bf n}_{x+\mu}-{\bf n}_{x}]
$.
ii) The  derivative (\ref{cderivative}) reproduces correctly the continuum covariant derivative for the adjoint field up to ${\cal O}(\epsilon)$:
\begin{align}
 \epsilon^{-1} [V_{x,\mu} {\bf n}_{x+\mu} - {\bf n}_{x} V_{x,\mu}] 
 =& \epsilon^{-1} [{\bf 1} -i\epsilon g\mathbf{V}_\mu(x) + {\cal O}(\epsilon^2)]{\bf n}_{x+\mu} - {\bf n}_{x} \epsilon^{-1} [{\bf 1} -i\epsilon g\mathbf{V}_\mu(x) + {\cal O}(\epsilon^2)] 
 \nonumber\\
=& \epsilon^{-1}[ {\bf n}_{x+\mu} - {\bf n}_{x}] - i g[\mathbf{V}_\mu(x) {\bf n}_{x+\mu} - {\bf n}_{x} \mathbf{V}_\mu(x)] + {\cal O}(\epsilon) ,
 \nonumber\\
=& \partial_\mu^{(\epsilon)} {\bf n}_{x} - i g[\mathbf{V}_\mu(x), {\bf n}_{x}] + {\cal O}(\epsilon) ,
\label{r1}
\end{align}
where we have used the  ${\cal O}(\epsilon^2)$ ambiguity in the last step.%
\footnote{
Adopting another form instead of (\ref{r1}) using the ambiguity of ${\cal O}(\epsilon^2)$, 
$
 0 = {\bf n}_{x+\mu} - {\bf n}_{x}  - i \epsilon g  \mathbf{V}_\mu(x)  {\bf n}_{x}
 +  i {\bf n}_{x+\mu}  \epsilon g  \mathbf{V}_\mu(x)   
   + {\cal O}(\epsilon^2),
$
which is rewritten as
$
  [ \bm{1}   +  i  \epsilon g  \mathbf{V}_\mu(x)  + {\cal O}(\epsilon^2) ]{\bf n}_{x}
 =  {\bf n}_{x+\mu} [ \bm{1} + i  \epsilon g  \mathbf{V}_\mu(x) + {\cal O}(\epsilon^2)] ,
$
we obtain a relation 
$
 V_{x,\mu}^\dagger  {\bf n}_{x}
 = {\bf n}_{x+\mu} V_{x,\mu}^\dagger  .
$
However, this is nothing but the Hermitian conjugate of  (\ref{Lcc}) and does not lead to a new condition. 
The definition of the covariant derivative  could be improved e.g., by using a symmetric difference, as will be discussed in a separate paper. 
}
iii) The derivative (\ref{cderivative}) obeys the correct transformation property, i.e., the adjoint rotation on a lattice: 
\begin{align}
  D_\mu^{(\epsilon)}[\mathbf{V}] {\bf n}_{x} \rightarrow \Omega_{x}(D_\mu^{(\epsilon)}[\mathbf{V}] {\bf n}_{x})\Omega_{x+\mu}^\dagger ,
\end{align}
provided that the link variable $V_{x,\mu}$ transforms in the same way as the original link variable $U_{x,\mu}$:
\begin{align}
  V_{x,\mu} \rightarrow \Omega_{x} V_{x,\mu} \Omega_{x+\mu}^\dagger
   = V_{x,\mu}' .
   \label{transf-V}
\end{align}
This is required from the transformation property of the continuum variable $\mathbf{V}_\mu(x)$, see \cite{KMS05}.
Therefore, we obtain the desired condition (\ref{Lcc}) between ${\bf n}_{x}$ and $V_{x,\mu}$.
The  defining equation (\ref{Lcc}) for the link variable $V_{x,\mu}$  is form-invariant under the gauge transformation II, i.e.,
$ {\bf n}_{x}^\prime V_{x,\mu}^\prime  = V_{x,\mu}^\prime {\bf n}_{x+\mu}^\prime
$. 

A lattice version of the orthogonality equation (\ref{nX=0}) is given by ${\rm tr}({\bf n}_{x} {\bf X}_\mu(x))=0$ or
\begin{equation}
 {\rm tr}({\bf n}_{x} \exp \{-i\epsilon g {\bf X}_\mu(x)\})
  =  {\rm tr}({\bf n}_{x}  \{ {\bf 1}-i\epsilon g {\bf X}_\mu(x) \} ) + {\cal O}(\epsilon^2) = 0 + {\cal O}(\epsilon^2) .
  \label{cond2}
\end{equation}
This implies that the trace vanishes  up to first order of $\epsilon$ apart from the second order term. 
Remembering  the relation ${\bf X}_\mu(x)={\bf A}_\mu(x)-{\bf V}_\mu(x)$, we can rewrite   (\ref{cond2}) into (\ref{cond2m}) in terms of ${\bf n}_{x}$ and $U_{x,\mu}$.
Note that the orthogonality condition (\ref{cond2m}) is gauge invariant. 

First, we proceed to solve the defining equation (\ref{Lcc}) for the link variable $V_{x,\mu}$ and express it in terms of the site variable ${\bf n}_{x}$ and the original link variable $U_{x,\mu} = \exp( -i \epsilon g \mathbf{A}_\mu(x))$, 
just as the continuum variable $\mathbf{V}_\mu(x)$ is expressed in terms of ${\bf n}(x)$ and $\mathbf{A}_\mu(x)$ in (\ref{Vdef}). 
The equation (\ref{Lcc}) is linear in $V_{x,\mu}$. Therefore, the normalization of $V_{x,\mu}$ can not be determined by this equation alone.
In general,  unitarity is not guaranteed for the general solution of the defining equation and hence a unitarity condition must be imposed afterwards. 
Moreover, the equation (\ref{Lcc}) is a matrix equation and it is rather difficult to obtain the general solution. 
Therefore, we adopt an ansatz (up to quadratic in ${\bf n}$): 
\begin{align}
 V_{x,\mu}
 = U_{x,\mu} + \alpha {\bf n}_{x} U_{x,\mu} + \beta U_{x,\mu} {\bf n}_{x+\mu} 
 + \gamma {\bf n}_{x} U_{x,\mu} {\bf n}_{x+\mu}  ,
 \label{ansatz}
\end{align}
which enjoys the correct transformation property, the adjoint rotation (\ref{transf-V}).
It turns out that this ansatz satisfy the defining equation,%
\footnote{
Uniqueness of the solution should be discussed separately.
}
if and only if the numerical coefficients $\alpha, \beta$ and $\gamma$ are chosen to be 
\begin{equation}
 \gamma= 1, \quad \alpha = \beta , 
\end{equation}
where 
we have used 
$
 {\bf n}_{x}{\bf n}_{x}=n_{x}^A n_{x}^B \sigma_A \sigma_B
=n_{x}^A n_{x}^B ( \delta_{AB} {\bf 1} + i \epsilon_{ABC} \sigma_C) 
=   \vec{n} \cdot \vec{n} {\bf 1} =   {\bf 1}
$.

Second, substituting the ansatz (\ref{ansatz}) with a still undetermined parameter $\alpha$ into the left-hand side of (\ref{cond2m}),  we obtain
\begin{equation}
 {\rm tr}({\bf n}_{x} U_{x,\mu} V_{x,\mu}^\dagger) 
  = \alpha^* [{\rm tr}({\bf 1}) + {\rm tr}({\bf n}_x U_{x,\mu} {\bf n}_{x+\hat{\mu}} U_{x,\mu}^\dagger)]
  =4  \alpha^*  
   + {\cal O}(\epsilon^2) ,
  \label{cond2mm}
\end{equation}
since
\begin{align}
&
{\rm tr}({\bf n}_x U_{x,\mu} {\bf n}_{x+\hat{\mu}} U_{x,\mu}^\dagger)
  ={\rm tr}
   \{{\bf n}_x
     (1-i\epsilon g{\bf A}_\mu^{(x)})
     ({\bf n}_x+\epsilon\partial_\mu^{(\epsilon)}{\bf n}_x)
     (1+i\epsilon g{\bf A}_\mu^{(x)})
     +{\cal O}(\epsilon^2)\}
   \nonumber\\
 =& {\rm tr}
   \{{\bf n}_x{\bf n}_x
     +i\epsilon g {\bf n}_x {\bf n}_x {\bf A}_\mu^{(x)}
     -i\epsilon g {\bf n}_x  {\bf A}_\mu^{(x)}{\bf n}_x
     +\epsilon{\bf n}_x\partial_\mu^{(\epsilon)}{\bf n}_x
     +{\cal O}(\epsilon^2)\}
 ={\rm tr}({\bf 1})
   +{\cal O}(\epsilon^2) 
\end{align}
where we have used 
$
 {\rm tr}({\bf n}_x  \partial_\mu^{(\epsilon)} {\bf n}_{x}  )
=0+ {\cal O}(\epsilon)
$. 
Hence, the condition (\ref{nX=0}) leads to $\alpha=0+ {\cal O}(\epsilon^2)$. 
Note that 
${\rm tr}({\bf n}_{x+\mu} V_{x,\mu}^\dagger U_{x,\mu} )=0$ gives the same condition as (\ref{cond2m})  by virtue of (\ref{Lcc}). 
[If we imposed  invariance on a lattice under the discrete global transformation, ${\bf n}_{x} \rightarrow - {\bf n}_{x}$ respected in the continuum theory, we would have obtained  $\alpha=0=\beta$.]
Thus we have determined $V_{x,\mu}$ 
up to an overall normalization constant 
\footnote{
This special form has already been invented in a different context in the paper \cite{CGI98} in order to give the gauge-invariant lattice definition of Nambu magnetic monopole with quantized magnetic charge  in the SU(2) Higgs model on a lattice, although we have given more general scheme to find such a form in this paper. 
}
\begin{align}
  V_{x,\mu} = V_{x,\mu}[U,{\bf n}] 
  = U_{x,\mu} +  {\bf n}_{x} U_{x,\mu} {\bf n}_{x+\mu} .
  \label{sol}
\end{align}

Even in this stage,   $V_{x,\mu}$ is not necessarily unitary, since 
 neither $V_{x,\mu}V_{x,\mu}^\dagger$ nor $V_{x,\mu}^\dagger V_{x,\mu}$   become a unit matrix ${\bf 1}$. In fact, we have
 \begin{align}
 V_{x,\mu}V_{x,\mu}^\dagger &=   2{\bf 1} + {\bf n}_x U_{x,\mu} {\bf n}_{x+\hat{\mu}} U_{x,\mu}^\dagger + U_{x,\mu} {\bf n}_{x+\hat{\mu}} U_{x,\mu}^\dagger {\bf n}_x  , 
 \nonumber\\
  V_{x,\mu}^\dagger V_{x,\mu} &=    2{\bf 1} + U_{x,\mu}^\dagger 
  {\bf n}_{x} U_{x,\mu}{\bf n}_{x+\mu} 
  + {\bf n}_{x+\mu} U_{x,\mu}^\dagger {\bf n}_{x} U_{x,\mu}   . 
 \label{VV}
\end{align}
For $V_{x,\mu}$ to become unitary merely by a normalization,  the right-hand sides of Eqs.(\ref{VV}) must be proportional to a unit matrix.   
We can show that  this is indeed the case. 
\footnote{
The parameterization, 
${\bf n}_{x}=n_{x}^i \sigma_i$ and 
$U_{x,\mu}=u_{x,\mu}^{0}+i u_{x,\mu}^{j} \sigma_j$,
leads to 
$
{\bf n}_x U_{x,\mu} {\bf n}_{x+\hat{\mu}} U_{x,\mu}^\dagger
=A_0 {\bf 1}+ i A_j \sigma_j
$
with {\it real} variables $A_0$ and $A_j$ ($j=1,2,3$):
$
 A_0 = (n_x^i u_{x,\mu}^i)(n_{x+\mu}^j u_{x,\mu}^j) + (u_{x,\mu}^0 n_{x}^k - \epsilon^{ijk}n_x^i u_{x,\mu}^j)(u_{x,\mu}^0 n_{x+\mu}^k + \epsilon^{\ell mk}n_{x+\mu}^\ell u_{x,\mu}^m) 
$. 
Then, using a fact that $U_{x,\mu} {\bf n}_{x+\hat{\mu}} U_{x,\mu}^\dagger {\bf n}_x$ and ${\bf n}_x U_{x,\mu} {\bf n}_{x+\hat{\mu}} U_{x,\mu}^\dagger$ are Hermitian conjugate to each other, we find 
$
 {\bf n}_x U_{x,\mu} {\bf n}_{x+\hat{\mu}} U_{x,\mu}^\dagger + U_{x,\mu} {\bf n}_{x+\hat{\mu}} U_{x,\mu}^\dagger {\bf n}_x 
 = 2 A_0 {\bf 1} .
$
For $\alpha=0$, therefore, we have 
$
 V_{x,\mu}V_{x,\mu}^\dagger=2(1+A_0){\bf 1}=V_{x,\mu}^\dagger V_{x,\mu}
$.
}
Thus the unitary link variable $\hat{V}_{x,\mu}[U,{\bf n}]$ is obtained after  the normalization: 
\begin{eqnarray} 
\hat{V}_{x,\mu} = 
\hat{V}_{x,\mu}[U,{\bf n}] := 
 V_{x,\mu}/\sqrt{\frac{1}{2}{\rm tr} [V_{x,\mu}^{\dagger}V_{x,\mu}]} .
\label{cfn-mono-4}
\end{eqnarray}
Moreover, it is shown in the Appendix that the naive continuum limit $\epsilon \rightarrow 0$ of the link variable (\ref{cfn-mono-4}) reduces to the continuum expression (\ref{Vdef}). 
Therefore, we can define the gauge-invariant flux, $\bar{\Theta}_P[U,{\bf n}]$,
(plaquette variable) by
\begin{eqnarray} 
\bar{\Theta}_{x,\mu\nu}[U,{\bf n}] := \epsilon^{-2}
{\rm arg} ( {\rm tr} \{({\bf 1}+{\bf n}_x)\hat{V}_{x,\mu}\hat{V}_{x+\hat{\mu},\nu}
\hat{V}_{x+\nu,\mu}^{\dagger}\hat{V}_{x,\nu}^{\dagger} \}/{\rm tr}({\bf 1})) .
\label{cfn-mono-5}
\end{eqnarray}
This definition (\ref{cfn-mono-5}) is SU(2) gauge invariant due to   cyclicity of the trace and the transformation property of the link variable and the site variable. 
This does not change  thanks to the defining equation, even if we insert the factor $({\bf 1}+{\bf n}_x)$ at different corners of a plaquette, e.g., 
$
\bar{\Theta}_{x,\mu\nu}[U,{\bf n}] \equiv \epsilon^{-2}
{\rm arg} ( {\rm tr} \{\hat{V}_{x,\mu}({\bf 1}+{\bf n}_{x+\hat{\mu}})\hat{V}_{x+\hat{\mu},\nu}
\hat{V}_{x+\nu,\mu}^{\dagger}\hat{V}_{x,\nu}^{\dagger} \})
$.

Finally,  we  show  quantization of the magnetic charge  as follows. 
We have constructed a color vector field  according to the adjoint orbit representation\cite{KKMSSI05}:
\begin{equation}
 {\bf n}_{x} = \Theta_{x} \sigma_3 \Theta_{x}^\dagger , \quad \Theta_{x} \in SU(2) , 
\end{equation}
which yields 
\begin{align}
  V_{x,\mu} 
  = U_{x,\mu} +  {\bf n}_{x} U_{x,\mu} {\bf n}_{x+\mu} 
  = \Theta_{x} [  \tilde{U}_{x,\mu} + \sigma_3  \tilde{U}_{x,\mu} \sigma_3] \Theta_{x+\mu}^\dagger , \quad 
   \tilde{U}_{x,\mu} :=  \Theta_{x}^\dagger U_{x,\mu} \Theta_{x+\mu} .
  \label{sol2}
\end{align}
By representing an SU(2) element $\tilde{U}_{x,\mu}$ in terms of Euler angles and Pauli matrices:
\begin{align}
   \tilde{U}_{x,\mu} :=  \Theta_{x}^\dagger U_{x,\mu} \Theta_{x+\mu} 
   = e^{i\sigma_3 \chi_{\ell}/2} e^{i \sigma_2 \theta_{\ell}/2} e^{i\sigma_3 \varphi_{\ell}/2} , \quad 
   \theta_\ell \in [0,\pi), \varphi_\ell, \chi_\ell \in [-\pi,\pi), 
\end{align}
the link variable $V_{x,\mu}$ for a link $\ell=(x,\mu)$ has the representation:
\begin{align}
  V_{x,\mu} =2 \cos \frac{\theta_{\ell}}{2}   \hat{V}_{x,\mu}, \quad
  \hat{V}_{x,\mu} = \Theta_{x}
  \begin{pmatrix} e^{i(\varphi_{\ell}+\chi_{\ell})/2} & 0 \\
  0 & e^{-i(\varphi_{\ell}+\chi_{\ell})/2} \\
  \end{pmatrix} \Theta_{x+\mu}^\dagger ,
\label{sol3}
\end{align}
with the normalization factor  given by 
$
 \sqrt{\frac{1}{2} {\rm tr}V_{x,\mu}V_{x,\mu}^\dagger} 
 = \sqrt{4 \cos^2 \frac{\theta_{\ell}}{2}}
 = 2 \cos \frac{\theta_{\ell}}{2}
$. 
Thus the gauge-invariant flux (\ref{cfn-mono-5}) is rewritten in terms of a compact variable $\Phi_\ell:=(\varphi_{\ell}+\chi_{\ell})/2 \in [-\pi,\pi)$ 
\begin{eqnarray} 
\bar{\Theta}_{x,\mu\nu}[U,{\bf n}] 
&:=& \epsilon^{-2}
{\rm arg} ( {\rm tr} \{({\bf 1}+{\bf n}_x)\hat{V}_{x,\mu}\hat{V}_{x+\hat{\mu},\nu}
\hat{V}_{x+\nu,\mu}^{\dagger}\hat{V}_{x,\nu}^{\dagger} \}/{\rm tr}({\bf 1})) 
\nonumber\\
&=& \epsilon^{-2}
{\rm arg} ( {\rm tr} \left\{ ({\bf 1}+\sigma_3) 
\begin{pmatrix} e^{i \sum_{\ell\in P}\Phi_{\ell}} & 0 \\
  0 & e^{-i\sum_{\ell\in P}\Phi_{\ell} } \\
  \end{pmatrix} 
\right\}/{\rm tr}({\bf 1})) 
\nonumber\\
&=&  \epsilon^{-2}
{\rm arg} \exp \{ i \Phi_{P} \} 
= [\Phi_{P}]_{{\rm mod}~2\pi}, 
\end{eqnarray}
where
\begin{equation}
 \Phi_{P} :=  (d\Phi)_{P} = \sum_{\ell\in P}\Phi_{\ell} 
 = \Phi_{x,\mu}+\Phi_{x+\mu,\nu}-\Phi_{x+\nu,\mu}-\Phi_{x,\nu} .
\end{equation}
It is important to remark that  $\bar{\Theta}_{x,\mu\nu}$ on a lattice is a compact variable whose range is $[-\pi,\pi)$, although it reduces to the continuum counterpart \cite{KKMSSI05} which is non-compact variable taking the value $(-\infty,\infty )$   in the continuum limit. 
This fact is crucial to  quantization of magnetic charge. 
In the unitary gauge, ${\bf n}_{x} \equiv \sigma_3$, 
which corresponds to $\Theta_{x}\equiv {\bf 1}$ in the above argument, 
$\bar{\Theta}_{x,\mu\nu}[U,{\bf n}]$ agrees with
$\bar{\theta}_{x,\mu\nu}$ in  the DT field strength
where the Abelian tensor $\theta_{\mu\nu}(s) \in [-4\pi,4\pi) \subset \mathbb{R}$ is decomposed into the field strength part $\bar{\theta}_{\mu\nu}(s) \in [-\pi,\pi) \subset \mathbb{R}$
and Dirac string part $n_{\mu\nu}(s)\in  \{-2,-1,0,1,2\} \subset \mathbb{Z}$:
$
 \bar{\theta}_{\mu\nu}(s)=\theta_{\mu\nu}(s)-2\pi n_{\mu\nu}(s) .
$
It is known \cite{CP97} that the elementary monopole defined in this way takes an integer-value $[-2,2]$, since the Bianchi identity holds for $\theta_{\mu\nu}(s)$.

\section{Numerical simulations}

As the algorithm for numerical simulations has already been given in a previous paper, we give a bit different explanation in this paper without repeating it. 
First of all, we generate the configurations of SU(2) link variables 
$\{ U_{x,\mu} \}$, $ U_{x,\mu}= \exp [ - ig\epsilon {\bf A}_\mu(x) ]$,
using the standard Wilson action based on the heat bath method. 
Next, we construct the color field variable $\mathbf{n}_{x}$ according to the following method. 
We minimize simultaneously the two functionals $F_{nMAG}$ and $F_{LLG}$ written in terms of the gauge (link) variable $U_{x,\mu}$ and the
color (site) variable $\mathbf{n}_{x}$:  
\begin{align}
 F_{nMAG}[U,\mathbf{n};\Omega,\Theta]
:=&  \sum_{x,\mu}\mathrm{tr}(\mathbf{1}-{}^{\Theta}\mathbf{n}_{x}{}^{\Omega}U_{x,\mu}{}^{\Theta}\mathbf{n}_{x+\mu}{}^{\Omega}U_{x,\mu}^{\dagger}),
\\
F_{LLG}[U;\Omega]
:=& \sum_{x,\mu}\mathrm{tr}(\mathbf{1}-{}^{\Omega} U_{x,\mu}) ,
\end{align}
with respect to two gauge transformations: 
${}^{\Omega}{}U_{x,\mu}:=\Omega_{x}U_{x,\mu}\Omega_{x+\mu}^{\dagger}$ for the
link variable $U_{x,\mu}$ and 
${}^{\Theta}\mathbf{n}_{x}:=\Theta_{x} \mathbf{n}_{x}^{(0)}\Theta_{x}^{\dagger}$ for an initial site variable
$\mathbf{n}_{x}^{(0)}$ (we can choose the initial value $\mathbf{n}_{x}^{(0)}=\sigma_{3}$)
where the gauge group elements $\Omega_{x}$ and $\Theta_{x}$
are {\it independent} SU(2) matrices on a site $x$. 
Then we can determine  the configurations  ${}^{\Theta^{*}}\mathbf{n}_{x}$ and ${}^{\Omega^{*}}U_{x,\mu}$ realizing the minimum of the first functional: $\min_{\Omega,\Theta}F_{nMAG}[U,\mathbf{n};\Omega,\Theta]=F_{nMAG}[U,\mathbf{n};\Omega^{*},\Theta^{*}]$, up to a common SU(2) transformation $G_{x}$.  
 This is because the \textquotedblleft common\textquotedblright  gauge transformation $G_{x}$ for $\Theta^{*}$ and $\Omega^{*}$ does not change the value of the functional $F_{nMAG}[U,\mathbf{n};\Omega,\Theta]$, i.e.,
$F_{nMAG}[U,\mathbf{n};\Omega^{*},\Theta^{*}]=F_{nMAG}[U,\mathbf{n};G\Omega^{*},G\Theta^{*}]$, since 
$
\mathrm{tr}({}^{\Theta}\mathbf{n}_{x}{}^{\Omega}U_{x,\mu}{}^{\Theta}\mathbf{n}_{x+\mu}{}^{\Omega}U_{x,\mu}^{\dagger})
$
$
= \mathrm{tr}(G_{x} {}^{\Theta}\mathbf{n}_{x} G_{x+\mu}^\dagger \cdot G_{x+\mu} {}^{\Omega}U_{x,\mu} G_{x+\mu}^\dagger  \cdot G_{x+\mu} {}^{\Theta}\mathbf{n}_{x+\mu} G_{x+\mu}^\dagger  \cdot G_{x+\mu} {}^{\Omega}U_{x,\mu}^{\dagger} G_{x}^\dagger)
$. 
This degrees of freedom for the SU(2) gauge transformation are fixed by minimizing the second functional  $F_{LLG}[U;\Omega]$ such that the configuration ${}^{\Omega^{**}}U_{x,\mu}$ realizes the minimum of the second functional: $\min_{\Omega}F_{LLG}[U;\Omega]=F_{LLG}[U;\Omega^{**}]$. 
Thus, imposing simultaneously two minimizing conditions removes the SU(2) ambiguity and the color field configuration are decided as 
$\hat{\mathbf{n}}_{x}:=\Theta_{x} \mathbf{n}_{x}^{(0)}\Theta_{x}^{\dagger}$ with
$\Theta_{x} = \Omega_{x}^{**}(\Omega_{x}^{*})^{-1}\Theta_{x}^{*} $ with $G_{x}=\Omega_{x}^{**}(\Omega_{x}^{*})^{-1}$.

Once  the configurations of the color vector field $\{{\bf n}_x\}$ are generated according to the method explained above together with the configurations of SU(2) link variables $\{U_{x,\mu}\}$, we can construct $\{\hat{V}_{x,\mu}[U,{\bf n}]\}$ from (\ref{cfn-mono-4}).
The numerical simulations are performed on an $8^4$ lattice   at $\beta=$2.2, 2.3, 2.35, 2.4, 
2.45, 2.5, 2.6 
and on $16^4$ lattice at $\beta=2.4$ by thermalizing 3000 sweeps respectively.

\subsection{Quantization of magnetic charges}

We construct the gauge-invariant field strength (\ref{cfn-mono-5}) to extract  configurations of the magnetic monopole current $\{k_{x,\mu}\}$  defined by
\begin{eqnarray}
k_{\mu}(s)=
-\frac{1}{4\pi}{\varepsilon}_{\mu\nu\rho\sigma}
\partial_{\nu}\bar{\Theta}_{\rho\sigma}(x+\mu)
\simeq
-\frac{1}{4\pi}{\varepsilon}_{\mu\nu\rho\sigma}
\partial_{\nu}G_{\rho\sigma}(x) .
\label{cfn-conti-20}
\end{eqnarray}
This definition agrees with our definition of the magnetic monopole in the continuum (divided by $2\pi$).
This definition of the monopole current should be compared with the conventional monopole current on a lattice defined according to DeGrand and Toussaint \cite{DT80} through  link variables on the dual lattice \cite{KKMSSI05}:%
\begin{eqnarray*}
  k_{\mu}(s)=\frac{1}{2}{\varepsilon}_{\mu\nu\rho\sigma}
                  \partial_{\nu}n_{\rho\sigma}(s+\mu)
                   = -\frac{1}{4\pi}{\varepsilon}_{\mu\nu\rho\sigma}
                  \partial_{\nu}\bar{\theta}_{\rho\sigma}(s+\mu) .
\end{eqnarray*}
The monopole current $k_{\mu}(s)$ defined in this way becomes an integer-valued variable, since integer-valued variables $n_{\rho\sigma}$ are used to count the number of Dirac strings going out through a plaquette. 

In our formulation, on the other hand, we have only the real variable $\bar{\Theta}_P[U,{\bf n}]$ at hand, and we are to calculate the monopole current using the final term in (\ref{cfn-conti-20}).  Therefore, it is not so trivial to obtain the integer-valued $k_{\mu}(s)$ from the real-valued $\bar{\Theta}_P[U,{\bf n}]$.  
To check quantization of the magnetic charge, we have made a histogram of
$
 K(s,\mu) := 2\pi k_\mu(s) = \frac{1}{2}{\varepsilon}_{\mu\nu\rho\sigma}
\partial_{\nu}\bar{\Theta}_{\rho\sigma}(x+\mu)
$,
 i.e., magnetic charge distribution. 
 Note that $K(s,\mu)$ 
  should become a multiple of $2\pi$ if the magnetic charge is quantized.  
Our numerical results show that $K(s,\mu)$ is completely separated into $0$ or $\pm 2\pi$ within an error of $10^{-10}$, see 
Table~\ref{table:hist-of-rmg}. 
We have checked that the data in  Table~\ref{table:hist-of-rmg}  exhaust in total all the configurations $N=4\times 8^4=16384$,
because the number $N_l$ of links in the $d$-dimensional lattice with a side length $L$ is given by  $N_l=dL^d$. 
This result clearly shows that the magnetic charge defined anew is quantized as expected from the general argument. 
We have observed that the conservation law of the monopole current holds, since the number of $+2\pi$ configurations is the same as that of $-2\pi$ configurations.
In contrast, 
Table~\ref{table:hist-of-rmg} shows that quantization does not occur for the (old) CFN monopole constructed in the previous paper \cite{KKMSSI05} using  $H_{\mu\nu}$ alone. 
 
\begin{table}[h]
\caption{Histogram  of the magnetic charge (value of $K(s,\mu)$)   distribution 
for  new and  old monopoles on  $8^4$ lattice at $\beta=2.35$. }
\label{table:hist-of-rmg}
\begin{center}
\begin{footnotesize}
\begin{tabular}{cll}\hline
Charge & Number (new) & Number (old)  \\ \hline
-7.5$\sim$-6.5 & 0   & 0 \\
-6.5$\sim$-5.5 & 299 & 0 \\
-5.5$\sim$-4.5 & 0   & 1 \\
-4.5$\sim$-3.5 & 0   & 19\\
-3.5$\sim$-2.5 & 0   & 52\\
-2.5$\sim$-1.5 & 0   & 149\\
-1.5$\sim$-0.5 & 0   & 1086  \\
-0.5$\sim$0.5  & 15786 & 13801 \\
0.5$\sim$1.5   & 0   & 1035\\
1.5$\sim$2.5   & 0   & 173\\
2.5$\sim$3.5   & 0   & 52\\
3.5$\sim$4.5   & 0   & 16\\
4.5$\sim$5.5   & 0   & 0\\
5.5$\sim$6.5   & 299 & 0 \\
6.5$\sim$7.5   & 0   & 0\\
\hline
\end{tabular}
\end{footnotesize}
\end{center}
\end{table}

\subsection{Magnetic-current density}

Storing 30 configurations of the monopole current, 
we have calculated the magnetic-current density $\rho_{mon}$ defined by
\begin{eqnarray} 
\rho_{mon}=  <\sum_{x,\mu} |k_{\mu}(x)|>/(4V) .
\label{simulation-5}
\end{eqnarray}
The results are summarized in 
the second column of 
Table~\ref{table:density-of-mg}.  This should be compared with the first column of the conventional DT magnetic-current density \cite{Sijs91} and the third column of the old (naive) CFN one calculated from $H_{\mu\nu}$ alone \cite{KKMSSI05}.  The DT magnetic-current density is calculated from the photon potential $A_\mu^3$ extracted by using the Abelian projection \cite{tHooft81} in the MA gauge. 
The new monopole agrees with DT  monopole in density to high accuracy better than the naive CFN monopole.  
The similarity of the magnetic charge and current-density distributions between the DT monopole and the new monopoles suggests the dominance of the new magnetic  monopole  in the string tension, as will be discussed shortly. 

Note that  the new monopole always keeps gauge invariance by including both the electric and magnetic contributions coming from $c_{\mu}$ and $H_{\mu\nu}$ respectively. 
In contrast to the naive CFN monopole, however, the new monopole is difficult to be separated into the electric and magnetic parts in this sense.

\begin{table}[h]
\caption{
$\beta$ dependence of magnetic-current densities 
for DT monopole, new and old CFN monopole. 
}
\label{table:density-of-mg}
\begin{center}
\begin{tabular}{clll}\hline
$\beta$ & DT monopole & new & old \\ \hline
2.2  & 0.0825(4)   & 0.0821(7)  &  0.0733(5)  \\ 
2.3  & 0.0515(3)   & 0.0514(12) &  0.0540(6)  \\
2.35 & 0.0379(3)   & 0.0371(7)  &  0.0423(8)  \\
2.4  & 0.0263(4)   & 0.0257(7)  &  0.0326(8)  \\
2.45 & 0.0167(6)   & 0.0172(9)  &  0.0233(10) \\
2.5  & 0.0096(5)   & 0.0108(8)  &             \\
2.6  & 0.0040(2)   & 0.0036(3)  &             \\
\hline
\end{tabular}
\end{center}
\end{table}

\subsection{Monopole dominance of the string tension}

In order to study  magnetic monopole dominance in the string tension, we proceed to estimate the magnetic monopole contribution $\left< W_m(C) \right>$ to the Wilson loop average $\left< W_f(C) \right>$, i.e., the expectation value of the Wilson loop operator.  
We define the magnetic part $W_m(C)$ of the Wilson loop operator $W_f(C)$ as the contribution from the monopole current $k_{\mu}(s)$ to the Wilson loop operator:
\footnote{
The Wilson loop operator $W_f(C)$ is decomposed into the magnetic part $W_m(C)$ and the electric part $W_e(C)$, which is derived from the non-Abelian Stokes theorem, see Appendix B of \cite{Kondo00}.
In this paper, we do not calculate the electric contribution $\left< W_e(C) \right>$ where $W_e(C)$ is expressed by the electric current $j_{\mu}=\partial_{\nu}F_{\mu\nu}$. 
}
\begin{align}
 W_m(C)     =&
     \exp\left\{
      2\pi i \sum_{s,\mu}k_{\mu}(s)N_{\mu}(s)
          \right\} ,
 \label{monopole dominance-1}     
\\
N_{\mu}(s)  =& \sum_{s'}\Delta_L^{-1}(s-s')\frac{1}{2}
\epsilon_{\mu\alpha\beta\gamma}\partial_{\alpha}
S^J_{\beta\gamma}(s'+\hat{\mu}), 
\quad
\partial'_{\beta}S^J_{\beta\gamma}(s) = J_{\gamma}(s) ,
\label{monopole dominance-2}
\end{align}
where $N_{\mu}(s)$ is defined through the external electric source $J_{\mu}(s)$ which is used to calculate the static potential:
$\partial'$ denotes the backward lattice derivative
$\partial_{\mu}^{'}f(x)=f(x)-f(x-\mu)$,  $S^J_{\beta\gamma}(s)$ denotes a surface bounded by the closed loop $C$ on which the electric source $J_{\mu}(s)$ has its support, and $\Delta_L^{-1}(s-s')$ is the Lattice Coulomb propagator. 
We obtain the string tension by evaluating the average of (\ref{monopole dominance-1}) from the generated configurations of the monopoles $\{k_{\mu}(s)\}$.


\begin{figure}[ptb]
\begin{center}
\vspace{-5mm}%
\includegraphics[width=3.0in]{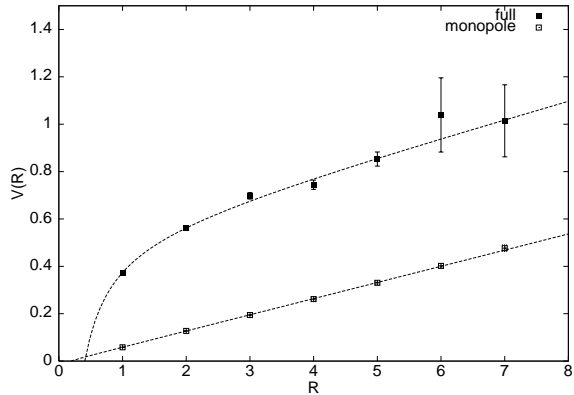}
\vspace{-8mm}
\end{center}
\caption{
The full SU(2) potential $V_f(R)$ and the magnetic--monopole potential $V_m(R)$ as functions of $R$ at $\beta=2.4$ on $16^4$ lattice.
\footnotesize 
}
\label{fig:potential}%
\end{figure}


The numerical simulations are performed on $8^4$ lattice at $\beta=2.3, 2.4$, and  $16^4$ lattice at $\beta=2.4$ by using  50 configurations in each case with 100 iterations.  
In particular, we have used 100 configurations for the calculation of the full string tension at $\beta=2.4$.
The Wilson loop average $\left< W(R,T)\right>$ for a rectangular loop $C=(R,T)$ with side lengths $R$ and $T$ is calculated by varying $R$ from $1$ to $4$ for a fixed $T$ ($T=4$) on $8^4$ lattice,  
and  from $1$ to $7$ for a fixed $T$ ($T=7$)  on $16^4$ lattice. 
Then we have calculated the respective potential $V_i(R)$ from the respective average $\left<W_i(C)\right>$:
\begin{eqnarray}
V_i(R) = -\log \left\{ \left< W_i(R,T) \right>/\left<W_i(R,T-1)\right> \right\} 
\quad (i=f, m) .
\label{monopole dominance-3}
\end{eqnarray}
Fig.~\ref{fig:potential} shows the full SU(2) potential $V_f(R)$ and the magnetic--monopole potential $V_m(R)$ as functions of $R$ at $\beta=2.4$ on $16^4$ lattice. 
\footnote{
It should be remarked that in Fig.~2 of Stack et al.\cite{SNW94}, a constant term was added to a magnetic-monopole potential to compare the $R$-dependence of the potential.
} 
Moreover, the numerical potential is $\chi^2$ fitted to the form with a linear term,  the Coulomb term and a constant term: 
\begin{eqnarray}
V_i(R) = \sigma_i R -  \alpha_i/R +c_i ,
\label{monopole dominance-4}
\end{eqnarray}
where $\sigma$ is the string tension, $\alpha$ is the Coulomb coefficient, 
and $c$ is the coefficient of the perimeter decay:
$\left<W_i(R,T)\right> \sim \exp [-\sigma_i RT -c_i(R+T)+\alpha_i T/R + \cdots]$. 
The results are shown in 
 Table~\ref{strint-tension-1}.

\begin{table}[h]
\caption{String tension and Coulomb coefficient I}
\label{strint-tension-1}
\begin{center}
\begin{tabular}{cllll}\hline
$\beta$ & $\sigma_f$ & $\alpha_f$ & $\sigma_m$ & $\alpha_m$ \\ \hline
2.3($8^4$)  & 0.158(14)   & 0.226(44)  &  0.135(13) & 0.009(36)  \\
2.4($8^4$)  & 0.065(13)   & 0.267(33)  &  0.040(12) & 0.030(34)  \\
{\bf 2.4($16^4$)} & 0.075(9)   & 0.23(2)  &  {\bf 0.068(2)}  & 0.001(5)\\
\hline
\end{tabular}
\end{center}
\end{table}

Fig.~\ref{fig:potential} and Table~\ref{strint-tension-1} show that the magnetic-monopole potential $V_m(R)$ has a dominant linear term and a negligibly small Coulomb term with small errors.  Consequently, the string tension $\sigma_m$ is obtained within small errors. 
In order to obtain the full SU(2) potential $V_f(R)$, on the other hand, various techniques have been used to eliminate as much as possible the unphysical short-distance fluctuations \cite{MPT95}. 
Especially, we have used  (APE's) smearing method \cite{albanese87} as a noise reduction technique.  
In this method, the Wilson loop average is calculated by using the block-link variable $U'_{x,\mu}$ constructed from the link variable $U_{x,\mu}$ according to the procedure:
\begin{eqnarray}
U'_{x,\mu} = U_{x,\mu} +\alpha \sum_{\nu\ne \mu \atop{\nu\ne 4}}
U_{x,\nu}U_{x+\hat{\nu},\mu}U^{\dagger}_{x+\hat{\mu},\nu} ,
\label{block link fields}
\end{eqnarray}
where $\alpha$ is a parameter to be adjusted later and the blocking is performed only for spacial directions, not for the time direction.  Of course, the obtained $U'_{x,\mu}$ is not unitary and hence it should be projected to be a unitary matrix $\hat{U}'_{x,\mu}$ by a standard method. 
This procedure is repeated  $N$ times iteratively starting from initial configurations obtained by the heat bath method. 
The parameters $N$ and $\alpha$ are chosen so that the Wilson loop average takes the maximal value.
In our calculations for the full SU(2) potential on $16^4$ lattice, we have used 100 configurations (twice the number of the other cases) to decrease statistical errors of $\sigma_f$ and $\alpha_f$, and adopted $N=15$ and $\alpha=0.2$ according to   \cite{Ito-PhD}.
The gauge invariance is preserved by this procedure, since the block-link variable $U'_{x,\mu}$ ($\hat{U}'_{x,\mu}$) has the same transformation property as the link variable $U_{x,\mu}$.

In fact, we find that 
the monopole part $\sigma_m$ reproduces 
85$\%$ of the full string tension $\sigma_f$ at $\beta=2.3$ on $8^4$ lattice  
and that $\sigma_m$ reproduces 91$\%$ of $\sigma_f$ at $\beta=2.4$ on $16^4$ lattice. 
Thus, we have confirmed the magnetic monopole dominance in the string tension using our magnetic
monopole in the gauge-invariant way.

\begin{table}[h]
\caption{String tension and Coulomb coefficient II (reproduced from \cite{SNW94})}
\label{strint-tension-2}
\begin{center}
\small
\begin{tabular}{cllll}\hline
$\beta$ & $\sigma_f$ & $\alpha_f$ & $\sigma_{DTm}$ & $\alpha_{DTm}$ \\ \hline
{\bf 2.4($16^4$)}  & 0.072(3)   & 0.28(2)  &  {\bf 0.068(2)} & 0.01(1)  \\
2.45($16^4$) & 0.049(1)   & 0.29(1)  &  0.051(1) & 0.02(1)  \\
2.5($16^4$)  & 0.033(2)   & 0.29(1)  &  0.034(1) & 0.01(1) \\
\hline
\end{tabular}
\end{center}
\end{table}

For comparison, we have shown in Table~\ref{strint-tension-2} the data of \cite{SNW94} which has discovered the monopole dominance for the first time on $16^4$ lattice where $\sigma_{DTm}$ reproduces 95$\%$ of $\sigma_f$. 
\footnote{
Stack et al. \cite{SNW94} have have performed numerical simulations using  500 configurations with 20 iterations on $16^4$ lattice in the calculation of the potential.  This will lead to less statistical errors for the full string tension by a factor $1/\sqrt{5}$ than ours, since we have used 100 configurations at $\beta=2.4$.  Indeed, our data are consistent with this estimation.
In addition to the smearing method, they used also the multihit method \cite{PPR83} to decrease noises in the calculation of the full string tension. 
Incidentally, they have measured the potential up to $R\times T=7\times 10$ on $16^4$ lattice. 
}
Here $\sigma_{DTm}$ and $\alpha_{DTm}$ denotes the conventional monopole contribution 
extracted from the diagonal potential $A_\mu^3$ using  Abelian projection in MAG. 
In particular, the comparison of the data on $16^4$ lattice at $\beta=2.4$ between Table~\ref{strint-tension-2} and  Table~\ref{strint-tension-1} reveals that the 
monopole contributions have exactly the same value between the conventional DT monopole and our
magnetic monopole.
This is because the monopole part does not include  the Coulomb term and hence the potential is obtained to an accuracy better than the full potential, as pointed above.

\section{Conclusion and discussion}

In this paper, we have proposed a new formulation of the NLCV of Yang-Mills theory which was once called the CFN decomposition.  This resolves all   drawbacks of the previous formulation \cite{KKMSSI05} on a lattice. 
This compact formulation  enables us to guarantee the magnetic charge quantization.
The new monopole dominance has been shown anew in the string tension. 
In this paper, the magnetic charge quantization and the magnetic monopole dominance in the string tension are confirmed in the gauge invariant way, whereas  
 they have been so far shown only in a special gauge fixing called  MA gauge which breaks the color symmetry explicitly.

A suitable definition of the link variable $\exp(-i\epsilon g {\bf X}_\mu(x))$ for  ${\bf X}_\mu(x)$ will be given with  relevant numerical results in a separate paper.  
Extending the promising formulation proposed here for SU(2)   to SU(3) will be given in a subsequent paper. 

\section*{Acknowledgments}
The authors would like to express their sincere thanks to  Maxim Chernodub for drawing their attention to the paper \cite{CGI98} and for very helpful discussions.
The numerical simulations have been done on a supercomputer (NEC SX-5) at  Research Center for Nuclear Physics (RCNP), Osaka University.
This project is also supported in part by the Large Scale Simulation Program No.133 (FY2005) of High Energy Accelerator Research Organization (KEK). 
This work is financially supported by 
Grant-in-Aid for Scientific Research (C)14540243 from 
JSPS
and in part by Grant-in-Aid for Scientific Research on Priority Areas (B)13135203 from MEXT.

\appendix
\section{Naive continuum limit}

We consider the naive continuum limit $\epsilon \rightarrow 0$ of the quantities on a lattice defined above  to see that they are good definitions on a lattice.

\subsection{Continuum limit of $V_{x,\mu}$}

The definitions of the gauge field $U_{x,\mu}=e^{-i\epsilon g \mathbf{A}_{\mu}(x)}$ ($\mathbf{A}_{\mu}(x)=\mathbf{A}^a_{\mu}(x) \frac{\sigma^a}{2}$) as a link variable 
and the adjoint scalar field ${\bf n}_x=n_x^a\sigma^a$ as a site variable  yield 
\begin{eqnarray} 
{\bf n}_xU_{x,\mu}{\bf n}_{x+\hat{\mu}}
&=& {\bf n}_x({\bf 1}-i\epsilon g\mathbf{A}_{\mu}(x)+{\cal O}(\epsilon^2))({\bf n}_x+\epsilon\partial_{\mu}{\bf n}_x+{\cal O}(\epsilon^2)) \nonumber \\
&=& {\bf n}_x^2 + \epsilon{\bf n}_x\partial_{\mu}{\bf }{\bf n}_x - i\epsilon g{\bf n}_x\mathbf{A}_{\mu}(x){\bf n}_x + {\cal O}(\epsilon^2) , 
\label{cfn-conti-1}
\end{eqnarray}
where we have used the relations: 
$(\sigma^a)^2 = {\bf 1}  (\text{no~summation})$,  
$\{\sigma^a,\sigma^b\}=2\delta^{ab}{\bf 1}$,   
$[\sigma^a,\sigma^b]=2i\epsilon^{abc}\sigma^c$,
which follow from 
$
\sigma^a\sigma^b= \delta^{ab} {\bf 1}+i\epsilon^{abc}\sigma^c .
$
The first term of (\ref{cfn-conti-1}) reads
$
{\bf n}_x^2 = n^a_xn^b_x\sigma^a\sigma^b = n^a_xn^b_x(\delta^{ab}{\bf 1}+i\epsilon^{abc}\sigma^c)
={\bf 1} ,
$
where we have used $(n^a)^2=n^a n^a=1$, $\epsilon^{abc}n^an^b=0$.
The second term of (\ref{cfn-conti-1}) reads 
$
{\bf n}_x\partial_{\mu}{\bf n}_x 
= n^a_x\partial_{\mu}n^b_x\sigma^a\sigma^b 
= -i\epsilon^{abc}(\partial_{\mu}n^a_x)n^b_x\sigma^c 
=: -2ig{\bf B}_{\mu}(x) ,
$
where we have used
$n^a_x\partial_{\mu}n^a_x=0$ which follows from differentiating $(n^a)^2=1$.
The third term of (\ref{cfn-conti-1}) reads 
\begin{eqnarray} 
i\epsilon g {\bf n}_x\mathbf{A}_{\mu}(x){\bf n}_x 
&=& i\epsilon g n^a_xA^b_{\mu}(x)n^c_x \sigma^a \sigma^b/2 \sigma^c \nonumber \\
&=& -i\epsilon g \mathbf{A}_{\mu}^a(x) \sigma^a/2 + i\epsilon g(n^a_x\mathbf{A}_{\mu}^a(x))n^c_x\sigma^c \nonumber \\
&=& -i\epsilon g \mathbf{A}_{\mu}(x) + 2i\epsilon g  c_{\mu}(x){\bf n}(x) ,
\label{cfn-conti-5} 
\end{eqnarray}
where we have used 
$.
$
twice and 
$\epsilon^{abd}\epsilon^{dce}=\delta^{ac}\delta^{be}-\delta^{ae}\delta^{bc}$, 
and  defined $c_{\mu}(x)$ by
$
c_{\mu}(x)\equiv n^a_x\mathbf{A}_{\mu}^a(x) .
$
Substituting these results into (\ref{cfn-conti-1}), we obtain 
\begin{eqnarray} 
{\bf n}_xU_{x,\mu}{\bf n}_{x+\hat{\mu}}
&=& {\bf 1} + i\epsilon g(\mathbf{A}_{\mu}(x) -2c_{\mu}(x){\bf n}(x) -2{\bf B}_{\mu}(x)) +{\cal O}(\epsilon^2)
\nonumber \\
&=& {\bf 1} + i\epsilon g(\mathbf{A}_{\mu}(x)-2\mathbf{V}_{\mu}(x) ) +{\cal O}(\epsilon^2) .
\label{cfn-conti-6} 
\end{eqnarray}
Therefore, we arrive at 
\begin{eqnarray} 
V_{x,\mu} &=& U_{x,\mu}+{\bf n}_xU_{x,\mu}{\bf n}_{x+\hat{\mu}} \nonumber \\
&=& ({\bf 1} - i\epsilon g \mathbf{A}_{\mu}(x))+({\bf 1}  + i\epsilon g(\mathbf{A}_{\mu}(x)-2\mathbf{V}_{\mu}(x) ) +{\cal O}(\epsilon^2) 
\nonumber \\
&=& 2({\bf 1}-i\epsilon g \mathbf{V}_{\mu}(x))+{\cal O}(\epsilon^2) = 2\hat{V}_{x,\mu}(U,{\bf n}) .
\label{cfn-conti-7} 
\end{eqnarray}
Hence, $V_{\mu}(x)$ in the continuum limit is calculated from 
$
\epsilon g {\bf V}_{\mu}(x) = -\frac{1}{2i}(\hat{V}_{x,\mu} - \hat{V}_{x,\mu}^{\dagger}) .
$

\subsection{SU(2) gauge-invariant field strength $\bar{\Theta}_{x,\mu\nu}$}

The plaquette variable $\hat{V}_{P}$ constructed from the link variable $\hat{V}_{x,\mu}$ reads 
\begin{equation}
\hat{V}_{P} := \hat{V}_{x,\mu}\hat{V}_{x+\hat{\mu},\nu}
\hat{V}_{x+\nu,\mu}^{\dagger}\hat{V}_{x,\nu}^{\dagger}
= \exp \{ {i\epsilon^2g\mathscr{F}_{\mu\nu}[\mathbf{V}]}  \} ,
\label{VVVV}
\end{equation}
with the field strength $\mathscr{F}_{\mu\nu}[\mathbf{V}]$ for $\mathbf{V}_\mu$ defined by 
$
 \mathscr{F}_{\mu\nu}[\mathbf{V}]=\partial_{\mu}\mathbf{V}_{\nu}-\partial_{\nu}\mathbf{V}_{\mu}
- ig [\mathbf{V}_{\mu},  \mathbf{V}_{\nu}] .
$
Then we have
\begin{eqnarray} 
{\rm tr} \{ {\bf n}_x \hat{V}_{P} \}
&=&
{\rm tr}({\bf n}_x) +i\epsilon^2g {\rm tr}({\bf n}_x \mathscr{F}_{\mu\nu}[\mathbf{V}]) + {\cal O}(\epsilon^4) \nonumber\\
&=&
i\epsilon^2g {\rm tr}\{
{\bf n}_x  (\partial_{\mu}\mathbf{V}_{\nu}-\partial_{\nu}\mathbf{V}_{\mu}
- ig [\mathbf{V}_{\mu},  \mathbf{V}_{\nu}])  \}+{\cal O}(\epsilon^4) \nonumber\\
&=&
i\epsilon^2g {\rm tr}({\bf 1}) \{
{\vec n} \cdot [\partial_{\mu}(c_{\nu}{\vec n}+ g^{-1} \partial_{\nu}{\vec n}\times {\vec n})
-\partial_{\nu}(c_{\mu}{\vec n}+g^{-1} \partial_{\mu}{\vec n}\times {\vec n}) \nonumber\\
&&
+g(c_{\mu}{\vec n}+g^{-1} \partial_{\mu}{\vec n}\times {\vec n})\times
(c_{\nu}{\vec n}+g^{-1} \partial_{\nu}{\vec n}\times {\vec n})
]\} +{\cal O}(\epsilon^4) \nonumber \\
&=&
i\epsilon^2 {\rm tr}({\bf 1}) \{
\partial_{\mu}c_{\nu}-\partial_{\nu}c_{\mu}
+g^{-1}  {\vec n}\cdot (\partial_{\mu}{\vec n}\times\partial_{\nu}{\vec n})
\}+{\cal O}(\epsilon^4) ,
\label{cfn-conti-17} 
\end{eqnarray}
where we have used 
$
 {\vec n}_x^2=1
$,
$
 (\vec{a}\times\vec{b})\times(\vec{a}\times\vec{c})=
(\vec{a}\cdot\vec{b}\times\vec{c})\vec{a}$
.
On the other hand, (\ref{VVVV}) leads to 
$
{\rm tr} \{ \hat{V}_{P} \}
 =  {\rm tr}({\bf 1}) + {\cal O}(\epsilon^4) .
$
Finally, we find that $\bar{\Theta}_{x,\mu\nu}$ has the same form as the 't Hooft tensor\cite{thooft74}:
\begin{eqnarray} 
{\rm tr} \{({\bf 1}+{\bf n}_x) \hat{V}_{P}  \}/{\rm tr}({\bf 1})
&=&
1+ i\epsilon^2(\partial_{\mu}c_{\nu}-\partial_{\nu}c_{\mu}
+ g^{-1} {\vec n}\cdot \partial_{\mu}{\vec n}\times\partial_{\nu}{\vec n})
\} \nonumber \\
&=&
\exp\{
i\epsilon^2(\partial_{\mu}c_{\nu}-\partial_{\nu}c_{\mu}
+ g^{-1}  {\vec n}\cdot \partial_{\mu}{\vec n}\times\partial_{\nu}{\vec n}
) \}
\nonumber\\
&=& \exp\{ i\epsilon^2\bar{\Theta}_{x,\mu\nu}\} .
\label{cfn-conti-19} 
\end{eqnarray}
Thus we have constructed a lattice version of the gauge-invariant field strength:
\begin{eqnarray} 
\bar{\Theta}_{x,\mu\nu} \simeq
\partial_{\mu}c_{\nu}-\partial_{\nu}c_{\mu}
+ g^{-1} {\vec n}\cdot (\partial_{\mu}{\vec n}\times\partial_{\nu}{\vec n})
\equiv G_{\mu\nu}(x) .
\label{cfn-fs} 
\end{eqnarray}



\end{document}